\documentclass[11pt,twoside]{article}
\usepackage{palatino}
\usepackage{graphicx}
\usepackage{amsmath, amssymb, amsbsy}
\usepackage[utf8]{inputenc}
\usepackage{t1enc}

\usepackage[mathcal]{eucal}

\usepackage{amsthm}

\usepackage[mathcal]{eucal}
\usepackage{array}

\def\contradict{\mathrel=\kern-2pt\mathrel|}

\usepackage[T1]{fontenc}

\usepackage{program}
\usepackage{alltt}
\usepackage{tabularx}

\newcommand{\Title}[1]{\setcounter{footnote}{0} {\begin{center}\Large\noindent \textbf{ #1}\end{center}}}
\newcommand{\Author}[1]{{\begin{center} \par\noindent\textbf{ #1}\end{center}}}
\newcommand{\keywords}[1]{
{\bigskip\par\noindent\small\textbf{Keywords:} #1}}

\newenvironment{Abstract}{\vspace*{0.em}\par\noindent\textbf{Abstract:~}}{}
\def\R{ \hbox{I\kern-.1667em R}}

\def\thebibliography#1{\subsubsection*{References}\list
 {[\arabic{enumi}]}{\settowidth\labelwidth{[#1]}\leftmargin\labelwidth
 \advance\leftmargin\labelsep
 \usecounter{enumi}}
 \def\newblock{\hskip .11em plus .33em minus .07em}
 \sloppy\clubpenalty4000\widowpenalty4000
 \sfcode`\.=1000\relax}

\def\Name1#1{{\vspace{0.5mm}\rm  #1:\par \vspace{1mm}}}

    {\begin{list}{\hbox{}}%
           {
       \setlength{\parsep}{0pt}
       \setlength{\itemsep}{0em}}}%
    {\end{list}}

\usepackage{subcaption}

\begin{document}

\Title{Corner-based implicit patches}
\Author{\'Agoston Sipos}

\begin{Abstract}
Multi-sided surfaces are often defined by side interpolants (also called ribbons), i.e. the surface has to connect to the ribbons with a prescribed degree of smoothness. The I-patch is such a family of implicit surfaces capable of interpolating an arbitrary number of ribbons and can be used in design and approximation. While in the case of parametric surfaces describing ribbons is a well-discussed problem, defining implicit ribbons is a different task.

This paper will introduce \textbf{corner I-patches}, a new representation that describes implicit surfaces based on corner interpolants. Those may be defined with much simpler surfaces, while the shape of the patch will depend on a handful of scalar parameters. Continuity between patches will be enforced via constraints on these parameters. Corner I-patches have several favorable properties that can be exploited for example in volume rendering or approximation.
\end{Abstract}

\section{Introduction}

Computer Aided Geometric Design deals with the mathematical representation of complex surface geometries. There is a wide variety of side interpolating multi-sided surfaces in the literature, including both parametric \cite{Charrot:1984,Varady:2011} and implicit \cite{Hartmann:2001,Varady:2001} patches. They are popular in curvenet-based design for their simple usage, as complex N-sided patches can be automatically created from simpler ribbon surfaces.

The common concept behind these patches is that \emph{ribbons} are introduced for each side, then \emph{blending functions}, that satisfy prescribed continuity constraints at the boundaries, mix those together. In the case of parametric multi-sided patches, ribbons usually are tensor-product surfaces. Blend functions are usually (not always) defined on a polygonal domain; surface points are calculated as a weighted sum of the evaluated points of the ribbons.

In the case of implicit surfaces, ribbons and blending functions are represented by implicit functions, defined on the whole 3D space. This, however, poses challenges, as an implicit ribbon, while interpolating the desired patch boundary, may have a very uneven shape inside the relevant space region; with sometimes disconnected branches or self-intersections.

For this reason, the polynomial degree of ribbons and blend functions is often desirable to be as low as possible. This, of course, poses limitations when used in design, as very detailed surfaces cannot be represented with a single patch. Locally defined surface elements are therefore important.

This paper explores the capabilities of a corner-based implicit formulation, where for each corner an interpolant has to be supplied, and the patch is created by a blend of them.

\section{Preliminaries}

Implicit surfaces are constant isosurfaces of real-valued functions on the 3D space. Usually the zero-isosurface is used: for a function $f : \mathbb{R}^3 \rightarrow \mathbb{R}$ the surface is $\{(x,y,z) \in \mathbb{R}^3 \, \mid \, f(x,y,z) = 0\}$.

Ribbon-based implicit surfaces are usually described in the following way. For each side, there is a given surface $R_i$, that the surface should smoothly connect to, with a given order of continuity. Then, there is a fixed equation of the patch, containing the $R_i$-s, and other defining surfaces.

\emph{Note:} in the following, capital letters in formulae will mean implicit functions ($\mathbb{R}^3 \rightarrow \mathbb{R}$), but the function arguments $(x,y,z)$ will be omitted for readability.

In the case of I-patches, which are the basis of the current research, the equation is
\begin{equation}
    {\displaystyle \sum_{i=1}^{n} \left( w_{i}R_{i}\prod_{\substack{j=1\\j\neq i}}^{n}B_{j}^{k}\right)+w_{0}\prod_{j=1}^{n}B_{j}^{k}}=0,
\end{equation}
where
\begin{itemize}
    \item $R_i$ are the ribbons, one for each side, to which the patch connects
    \item $B_i$ are the \emph{bounding surfaces}, whose intersection curves with the corresponding $R_i$ define the boundaries of the patch
    \item $0 \neq w_i \in \mathbb{R}$ are scalar parameters and $2 \leq k \in \mathbb{N}$ is an integer parameter
\end{itemize}

Consequently, the patch connects with $G^{k-1}$ connectivity to the ribbons along the bounding surfaces, as shown in \cite{Varady:2001}. It has also been proven that the I-patch is a consistent distance function with an inside and outside, in case of well-chosen signs of $w_i$-s \cite{Sipos:2020}.

In the following, we will use $k=2$ to keep the polynomial degree as low as possible, and in this paper, we discuss patches with $G^1$ continuity.

\section{Corner I-patch}

\subsection{Basic equation}

A corner I-patch is composed of the corner interpolants $S_{1,2}, S_{2,3}, ..., S_{n,1}$ and the bounding surfaces $B_1, B_2, ..., B_n$ (all of them non-coinciding implicit surfaces), such that $S_{i,i+1}$ is the corner between the $i$th and the $(i+1)$th boundaries.

Then, the equation of the corner I-patch is
\begin{equation}
\label{eq:cornerpatch}
    \displaystyle\sum_{i=1}^n \left(w_{i,i+1} \cdot S_{i,i+1} \cdot \prod_{\underset{j \neq i, j\neq i+1}{j=1}}^{n} B_j^2 \right) + \sum_{i=1}^{n} \left(w_i \cdot \prod_{\underset{j \neq i}{j=1}}^{n} B_j^2\right) + w \prod_{i=1}^n B_i^2,
\end{equation}
where the $w_{i,i+1}$ scalars can be merged into $S_{i,i+1}$, as multiplying with a nonzero number does not change the implicit isosurface, only its distance metric. Some important properties of this representation are:

\begin{itemize}
    \item In each corner, the patch connects with $G^1$ continuity to the corner interpolants. (This means that the gradient vectors of the surface have the same direction as the gradients of the interpolants there.)
    \item Along the $i$th boundary, the shape of the surface does not depend on $w$ and $w_j$ for $j \neq i$.
\end{itemize}

\subsection{Comparison to (side-based) I-patches}

A disadvantage of I-patches is that their gradient is a zero vector in the corner points. This can lead to unpleasant surfaces, thus it is generally better to be avoided. However, the gradient of the corner-I-patch can easily be proven to be the gradient of the corner interpolant times a nonzero number.

The corner-I-patch along the $i$th boundary connects smoothly to the implicit surface

\begin{equation}
\label{eq:side}
	S_{i-1,i} \cdot B_{i+1}^2 + S_{i,i+1} \cdot B_{i-1}^2 + w_i \cdot B_{i-1}^2 \cdot B_{i+1}^2.
\end{equation}

This itself is a 2-sided I-patch. Corner I-patches are thus similar - but not equivalent - to I-patches defined by ribbons that are themselves I-patches. (Such surfaces were described in \cite{Sipos:2020}.) This is because the I-patch defined by the ribbons in Equation \ref{eq:side} would be
\begin{equation}
\displaystyle\sum_{i=1}^n \left( S_{i,i+1} (B_{i+1}^2 B_{i-1}^2 + B_i^2 B_{i-2}^2) \prod_{\underset{j \neq i, j\neq i+1}{j=1}}^{n} B_j^2 \right) + \sum_{i=1}^{n} \left( w_i B_{i-1}^2 B_{i+1}^2 \prod_{\underset{j \neq i}{j=1}}^{n} B_j^2 \right) + w \prod_{i=1}^n B_i^2,
\end{equation}
which is not equivalent to Equation \ref{eq:cornerpatch}. Indeed, the factor $(B_{i+1}^2 B_{i-1}^2 + B_i^2 B_{i-2}^2)$ is what causes the I-patch's gradient to be zero at the corner points. This also causes corner-I-patches to have a lower degree of $2n$, as opposed to the $2n+2$ for these I-patches.

\subsection{Setting coefficients}

The $w_i$ and $w$ parameters can be set in a process similar to I-patches \cite{Varady:2001} forcing the patch to interpolate one point on each boundary and one point in the interior of the patch. As the surface shape on the $i$th boundary only depends on $w_i$, each of those can be set separately, and finally, $w$ can be set to interpolate the interior point. See examples in Figure \ref{fig:patchessame}.

\subsection{Limitations}

When connecting neighboring patches with geometric continuity, we need them to coincide at their common boundary in both a positional and a differential sense. As the corner-I-patch along $B_i$ connects to the surface defined by Equation \ref{eq:side}, the patch on the other side of $B_i$ also has to connect to it. This, however, only happens if $B_{i-1}$ and $B_{i+1}$ are also the corresponding bounding surfaces for the other patch.

This is not easily fulfilled when creating a general topology patchwork, but it is straightforward if the space is subdivided by planes, creating finite volume cells. Any such cell structure could theoretically work with corner-I-patches, however, the most practical and useful is to use a regular grid of cubes.

\section{Use in cell structures}

\begin{figure}[b]
     \centering
     \begin{subfigure}[b]{0.3\textwidth}
         \centering
         \includegraphics[width=\textwidth]{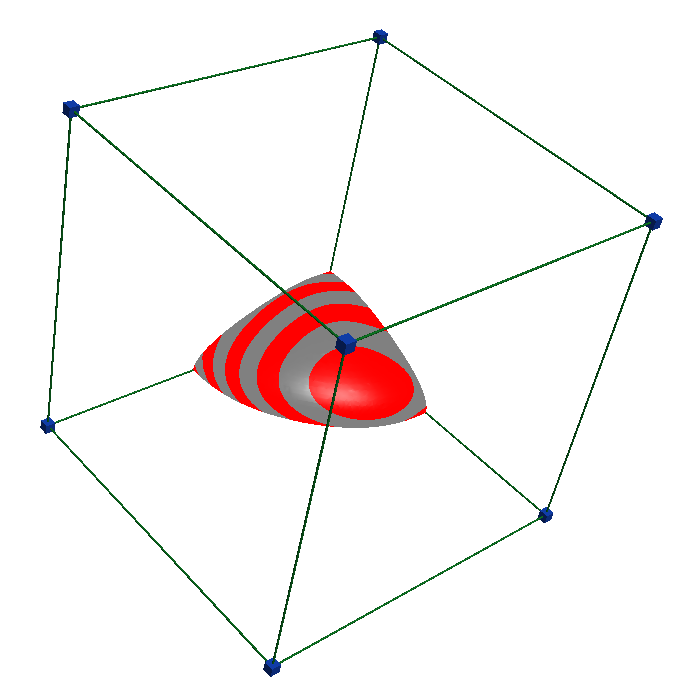}
         \caption{A 3-sided patch}
         \label{fig:patch1}
     \end{subfigure}
     \hfill
     \begin{subfigure}[b]{0.3\textwidth}
         \centering
         \includegraphics[width=\textwidth]{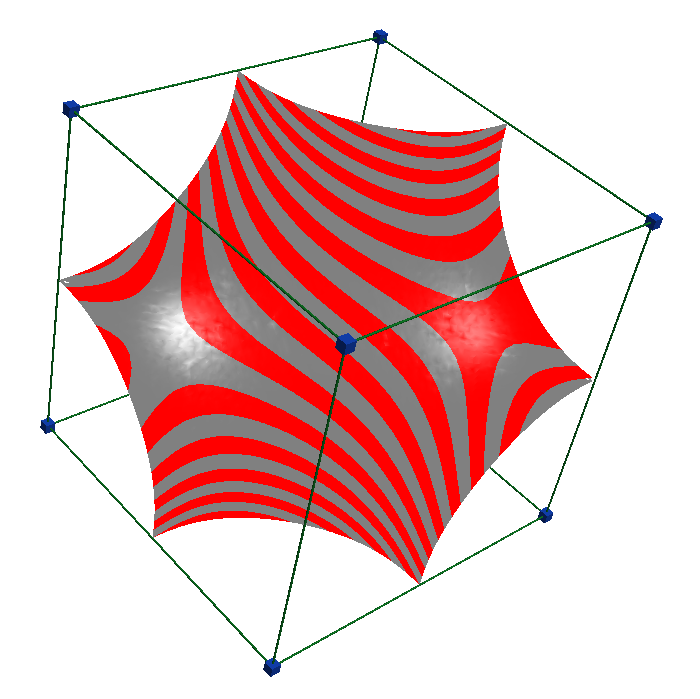}
         \caption{A 6-sided patch}
         \label{fig:patch2}
     \end{subfigure}
     \hfill
     \begin{subfigure}[b]{0.3\textwidth}
         \centering
         \includegraphics[width=\textwidth]{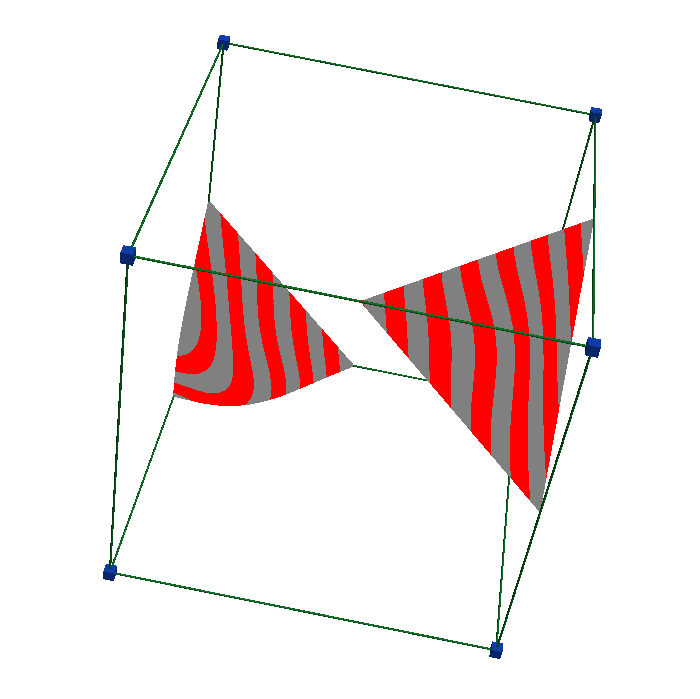}
         \caption{A patch consisting of two disjoint components}
         \label{fig:patch3}
     \end{subfigure}
        \caption{Corner I-patches inside the unit cube}
        \label{fig:patches}
\end{figure}

When used in regular cell structures, the $S_{i,j}$ and $B_i$ surfaces are all planes. Thus, the patch itself is a polynomial surface, with a degree of twice the number of sides (see Equation \ref{eq:cornerpatch}).

\begin{figure}[t]
     \centering
     \begin{subfigure}[b]{0.3\textwidth}
         \centering
         \includegraphics[width=\textwidth]{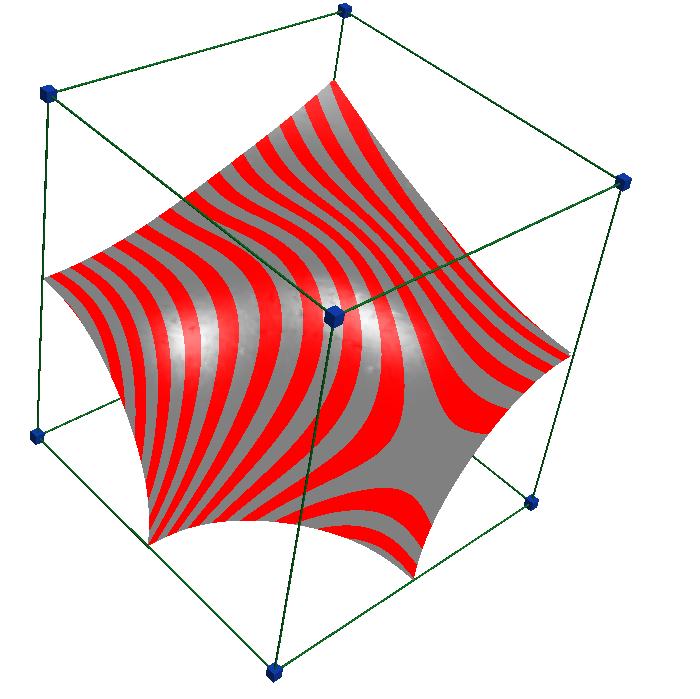}
         \caption{Base patch}
     \end{subfigure}
     \hfill
     \begin{subfigure}[b]{0.3\textwidth}
         \centering
         \includegraphics[width=\textwidth]{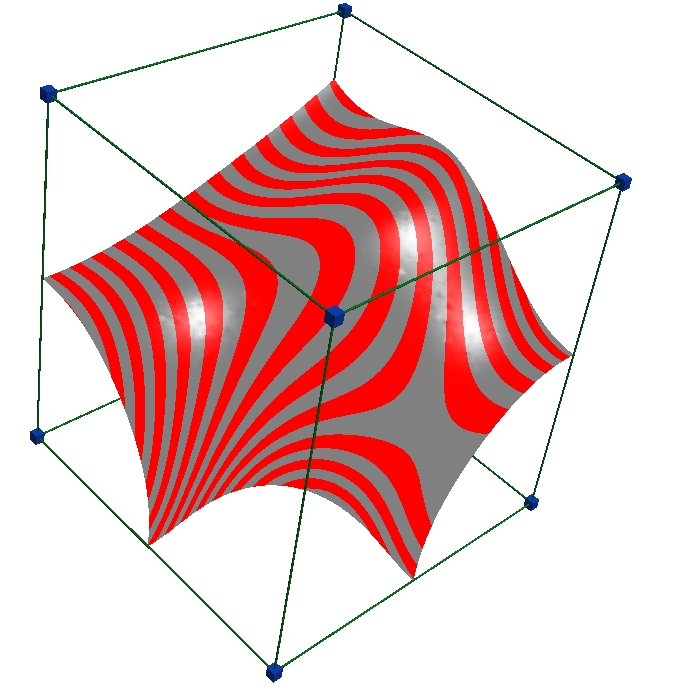}
         \caption{Changing boundaries}
     \end{subfigure}
     \hfill
     \begin{subfigure}[b]{0.3\textwidth}
         \centering
         \includegraphics[width=\textwidth]{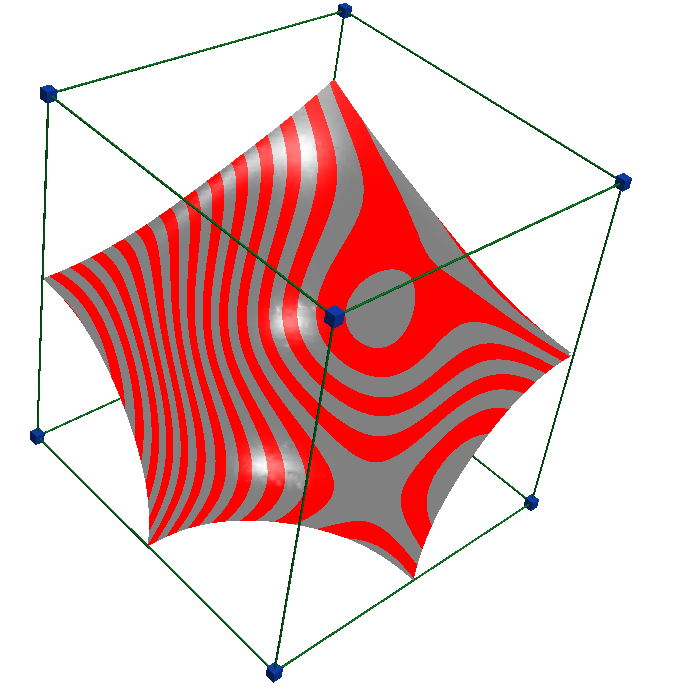}
         \caption{Changing interior}
     \end{subfigure}
        \caption{Corner I-patches with the same corners and different coefficients}
        \label{fig:patchessame}
\end{figure}

In Figures \ref{fig:patches} and \ref{fig:patchessame} corner patches are defined inside the unit cube. Figure \ref{fig:patch1} is a 3-sided surface near a corner of the cube. Figure \ref{fig:patch2} is a 6-sided patch that intersects all faces of the cube. In Figure \ref{fig:patchessame}, three patches with the same corners but different coefficients can be seen.

The possible topological configurations are similar to those of Marching Cubes \cite{lorensen1987marching}. With modified indexing, the scheme can also work for topologically disjoint isosurfaces (Figure \ref{fig:patch3}).

\section{Conclusion}

We have presented corner-I-patches, a class of implicit surfaces with several advantages over existing representations. Complex surfaces can be defined combining only planes, in contrast to the relatively more complicated ribbons needed to define I-patches. Corner-I-patches also get rid of the unwanted singularity in the corners. They have several scalar coefficients that can be used to optimize a target function on the patch for approximation or fairing purposes.

\section*{Acknowledgements}

This work was supported by the Hungarian Scientific Research Fund (OTKA, No. 124727). The author thanks his advisor Tamás Várady for helping to create the paper.

\bibliography{cikkek}
\bibliographystyle{plain}

\end{document}